\documentclass[%
aip,
rsi,%
amsmath,amssymb,
reprint,%
]{revtex4-1}

\usepackage{graphicx}% Include figure files
\usepackage{dcolumn}% Align table columns on decimal point
\usepackage{bm}% bold math
\usepackage{algorithm}%
\usepackage{algorithmicx}%
\usepackage{algpseudocode}%
\usepackage{float}
\usepackage[colorlinks=true, linkcolor=blue, citecolor=blue, urlcolor=blue]{hyperref}

\begin{document}
	
	\title{3D Photonic integration leveraging hybrid-confinement circuits}
	
	\author{Kanhaya Sharma}
	\affiliation{Universit\'{e} Marie et Louis Pasteur, CNRS UMR 6174, Institut FEMTO-ST, 15B Avenue Montboucons, Besançon, 25000, France}
	\email{kanhaya.sharma@femto-st.fr}	

	\author{Adri\`{a} Grabulosa}
	\affiliation{Universit\'{e} Marie et Louis Pasteur, CNRS UMR 6174, Institut FEMTO-ST, 15B Avenue Montboucons, Besançon, 25000, France}

	\author{Erik Jung}
	\affiliation{Currently with Heidelberg University, Kirchhoff Institute for Physics, Im Neuenheimer Feld 227, Heidelberg, Germany}

	\author{Daniel Brunner}
	\affiliation{Universit\'{e} Marie et Louis Pasteur, CNRS UMR 6174, Institut FEMTO-ST, 15B Avenue Montboucons, Besançon, 25000, France}
	
	\date{\today}% It is always \today, today,
	
	\begin{abstract}
		
Three-dimensional (3D) photonic integration offers a pathway to overcome the fundamental scaling limitations of planar platforms by enabling enhanced routing flexibility for compact, low-loss, and highly interconnected photonic circuits. 
 In this work, we fabricate 3D photonic circuits combining high-confinement air-clad waveguides for compact routing with low-confinement polymer-clad waveguides for robust single-mode operation within a monolithic platform.
 Efficient mode transition between polymer-clad and air-clad waveguides is demonstrated with a loss of 0.25~dB per interface.
 We also realize compact, Euler S- and U-shaped bends with minimal bending radii of 10~$\mu$m and losses as low as 0.5~dB and 0.4~dB, respectively, along with compact adiabatic air-clad splitters exhibiting a splitting loss of 0.6~dB over a length of 52~$\mu$m.
 Finally, full fabrication of a compact hybrid circuit is demonstrated, highlighting the feasibility and scalability of the approach.
 Our work represents a significant step in 3D photonic integration for applications including optical neural networks, photonic wire bonding and their potential for novel integrated photonic applications.
		
	\end{abstract}
	
	\maketitle
	
\section{Introduction}

Over the past decades, integrated electronic circuits have evolved from a few transistors to billions per chip, following the exponential scaling predicted by Moore’s law \cite{Moore_65, Clarke2005}. As this scaling approaches fundamental physical limits, conventional two-dimensional (2D) integration strategies are becoming increasingly constrained. Consequently, three-dimensional (3D) integration has emerged as a promising pathway to sustain performance scaling, driven by advances in 3D transistors, NAND architectures, and integrated circuit technologies \cite{WANG2023170, Ishihara2012, Colinge2013, Kelion2012}. Notably, industry developments such as multi-layer V-NAND flash memory have demonstrated vertically stacked architectures exceeding 400 layers in recent generations \cite{Shilov, Samsung}. However, such implementations fundamentally rely on planar fabrication processes, where 3D functionality is realized through repeated 2D patterning steps, leading to increased fabrication complexity and cost. In particular, the reliance on high-resolution lithographic masks restricts scalability to structures with a high degree of layer-to-layer similarity \cite{Shilov, Samsung}. 

3D integration has emerged not only as a viable approach in electronics but is becoming increasingly indispensable in photonics. Established 2D photonic platforms, including silicon photonics, silicon nitride, and lithium niobate widely used for data center interconnects, quantum communication, and optical computing \cite{Butt2025}, have also started following the layer stacking approach in order to realize integration into 3D photonic architectures \cite{Daudlin2025, Xiang2023}. For instance, \cite{Daudlin2025} has demonstrated high-bandwidth interchip data links using vertically stacked 3D integration of multiple optical transceivers, achieving an aggregate bandwidth approaching $800~\textrm{Gbs}^{-1}$ within a compact footprint of $0.3~\textrm{mm}^2$ and hence a bandwidth density of $5.3~\textrm{Tbs}^{-1}\textrm{mm}^{-2}$. In comparison, conventional planar 2D integrated photonic transceivers typically exhibit bandwidth densities on the order of $0.1$–$2~\textrm{Tbs}^{-1}\textrm{mm}^{-2}$ \cite{Zhou2024, Rizzo2023}. Furthermore, 3D monolithic heterogeneous integration has been demonstrated by bonding a III–V gain medium onto an ultra-low-loss silicon nitride platform, thereby merging laser operation and low-loss propagation at $0.5~\textrm{dB/m}$ on a single platform \cite{Xiang2023}. Extending such natively 2D fabrication layers in 3D through iterative stacking of planar layers introduces increased fabrication complexity, cost, and processing time, while also suffering from limited vertical resolution. These limitations pose significant challenges for emerging applications in optics such as 3D optical neural networks \cite{Moughames2020_spl}, photonic lanterns \cite{Dana2024}, and optical splitters requiring precise optical path-length control \cite{Grabulosa2023}.

\begin{figure*}[ht]
\centering\includegraphics[width=0.8\linewidth]{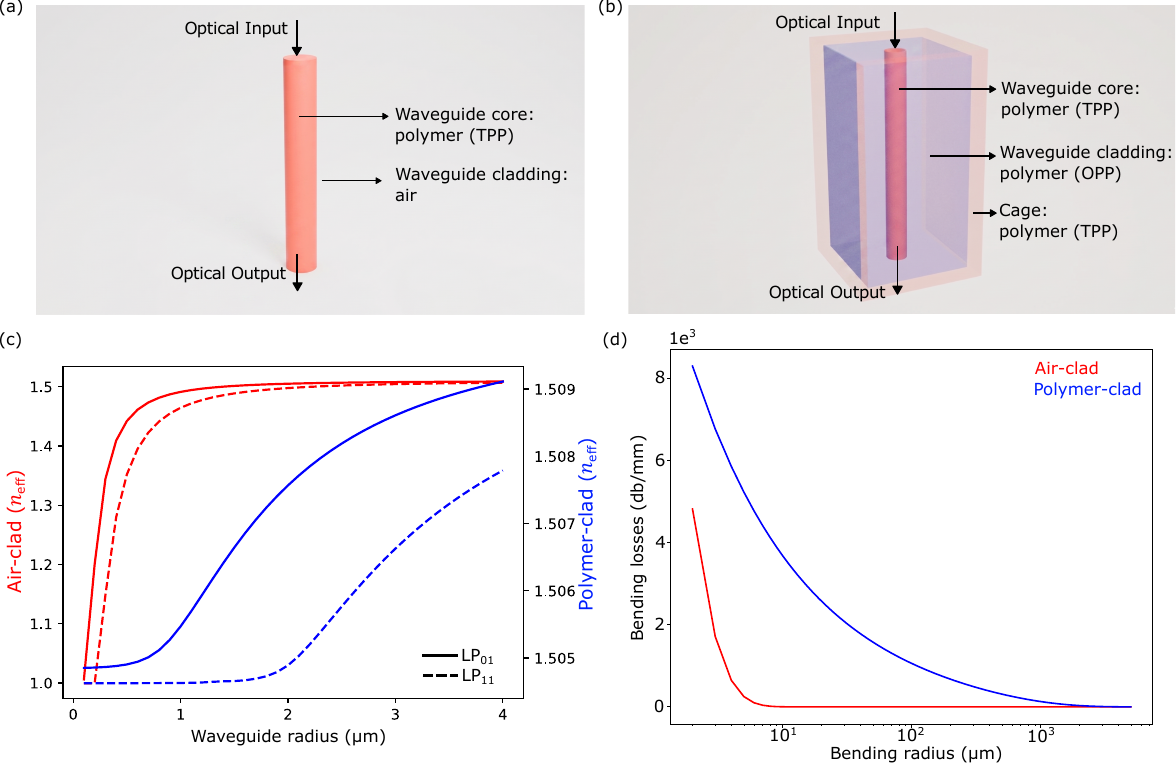}
\caption{(a) Schematic of an air-clad waveguide. (b) Schematic of a polymer-clad waveguide. (c) Effective index for the $\mathrm{LP}_{01}$ and $\mathrm{LP}_{11}$ modes of air-clad (red) and polymer-clad (blue) waveguides as a function of waveguide radius. (d) Bending losses of air-clad (red) and polymer-clad (blue) waveguides as a function of circular bend radius.}
\label{fig:figure1}
\end{figure*}

One viable and native 3D lithography approach to address the constraints of 2D is freeform 3D direct laser writing via two-photon polymerization (DLW-TPP), enabling true 3D integration of complex topologies. Such two-photon polymerization (TPP) allows high-resolution fabrication at sub-micrometer scales via true 3D maskless freeform writing combined with single-voxel level control of the refractive index \cite{Moughames2020, Porte2021, Grabulosa2022}. In addition, its non-contact fabrication process enables smooth surface morphologies, with surface roughness below 5~nm, and reduced post-processing complexity compared to other additive manufacturing techniques, including digital light processing and stereolithography, which generally require additional curing and surface finishing steps \cite{Xin2023, Camposeo2019, Chaudhary2023, Jadhav2022}. Leveraging these advantages allowed to realize densely interconnected 3D optical neural networks (ONNs), reducing footprint scaling from quadratic in 2D CMOS implementations to linear in additively manufactured 3D architectures \cite{Moughames2020}. Finally, the entire process has been demonstrated to be CMOS compatible \cite{Grabulosa2023, Billah2015, Huang2024}. 

In this work, we introduce and experimentally demonstrate the key building blocks for hybrid confinement 3D photonic waveguide networks including compact bends, splitters, and low-loss confinement transitions. The proposed platform enables single-mode propagation across photonic circuits combining low-confinement single-mode polymer-clad waveguides with highly confined multi-mode air-clad waveguides. By leveraging the advantages of both confinement regimes, the architecture combines precise mode control with dense photonic integration, enabling compact routing and tight bending geometries. This is achieved by tailoring the waveguide core radii and refractive index contrast to ensure the fundamental modes in both sections are matched, thereby minimizing excitation of higher-order modes in multi-mode air-clad waveguide. We experimentally determine losses as low as 0.25 dB per confinement-interface transition. Furthermore, single-mode propagation is maintained even for tight bending radii approaching 10 $\mu$m. To fabricate our circuits, we used a commercially available DLW-TPP system from Nanoscribe GmbH (Photonic Professional GT+) using IP-Dip photoresist with a 63$\times$ microscope objective for the individual study of bends and splitters, and IP-S photoresist with a 25$\times$ microscope objective for complete hybrid confinement circuit implementation. All structures in the manuscript are designed and characterized at a wavelength of 650 nm.

\begin{figure}[ht]
\centering\includegraphics[width=\linewidth]{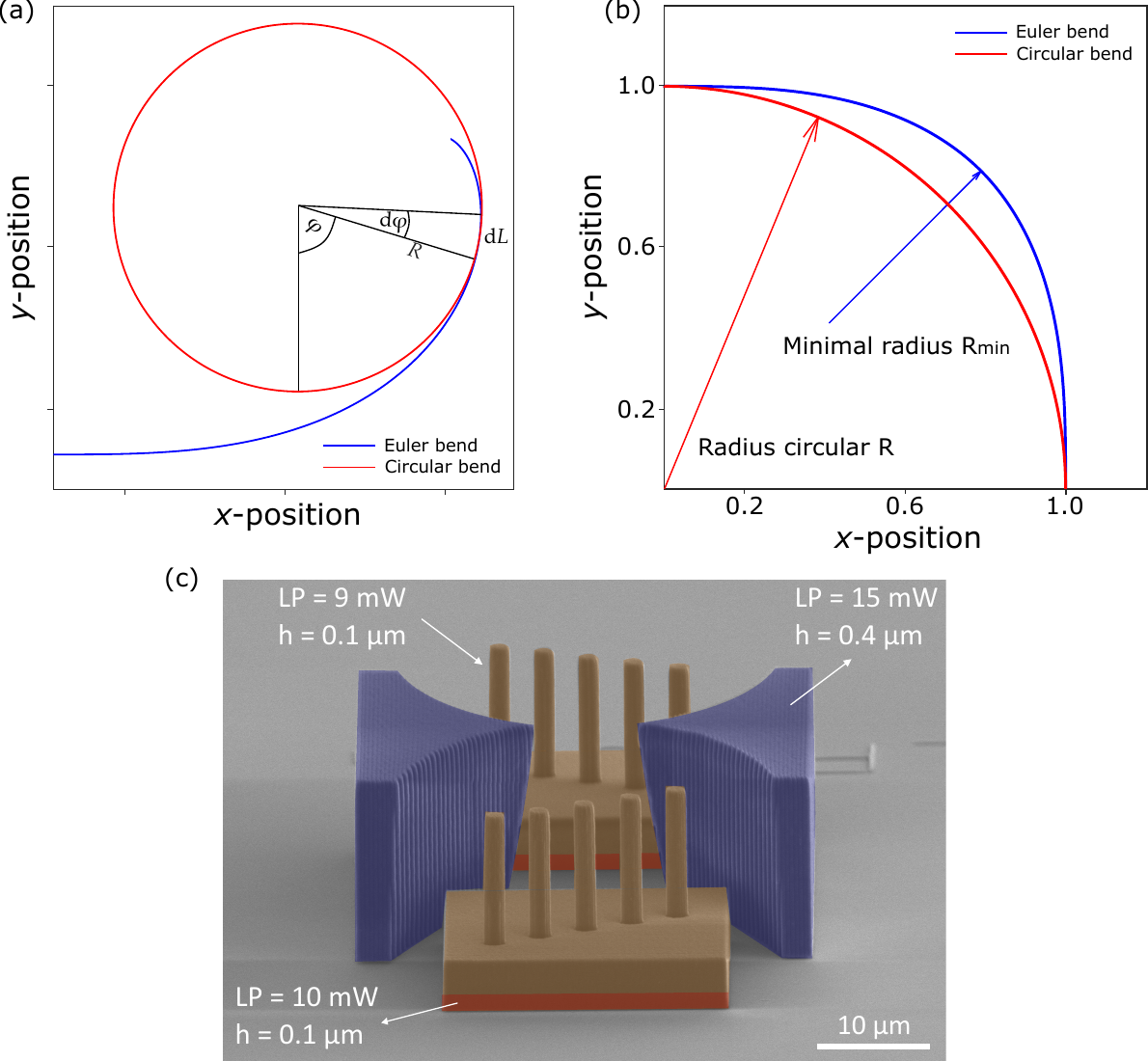}
\caption{(a) Approximation
of a clothoid for an infinitesimally small angle $d\phi$ through a circle with the same radius. (b) Comparison of the shape of $90^\circ$ Euler and circular bends. (c) SEM micrograph of a 3D printed S-bends with mechanical support (purple), bends (brown) and pedestal (orange).}
\label{fig:figure2}
\end{figure}

\begin{figure*}[ht]
\centering\includegraphics[width=0.7\linewidth]{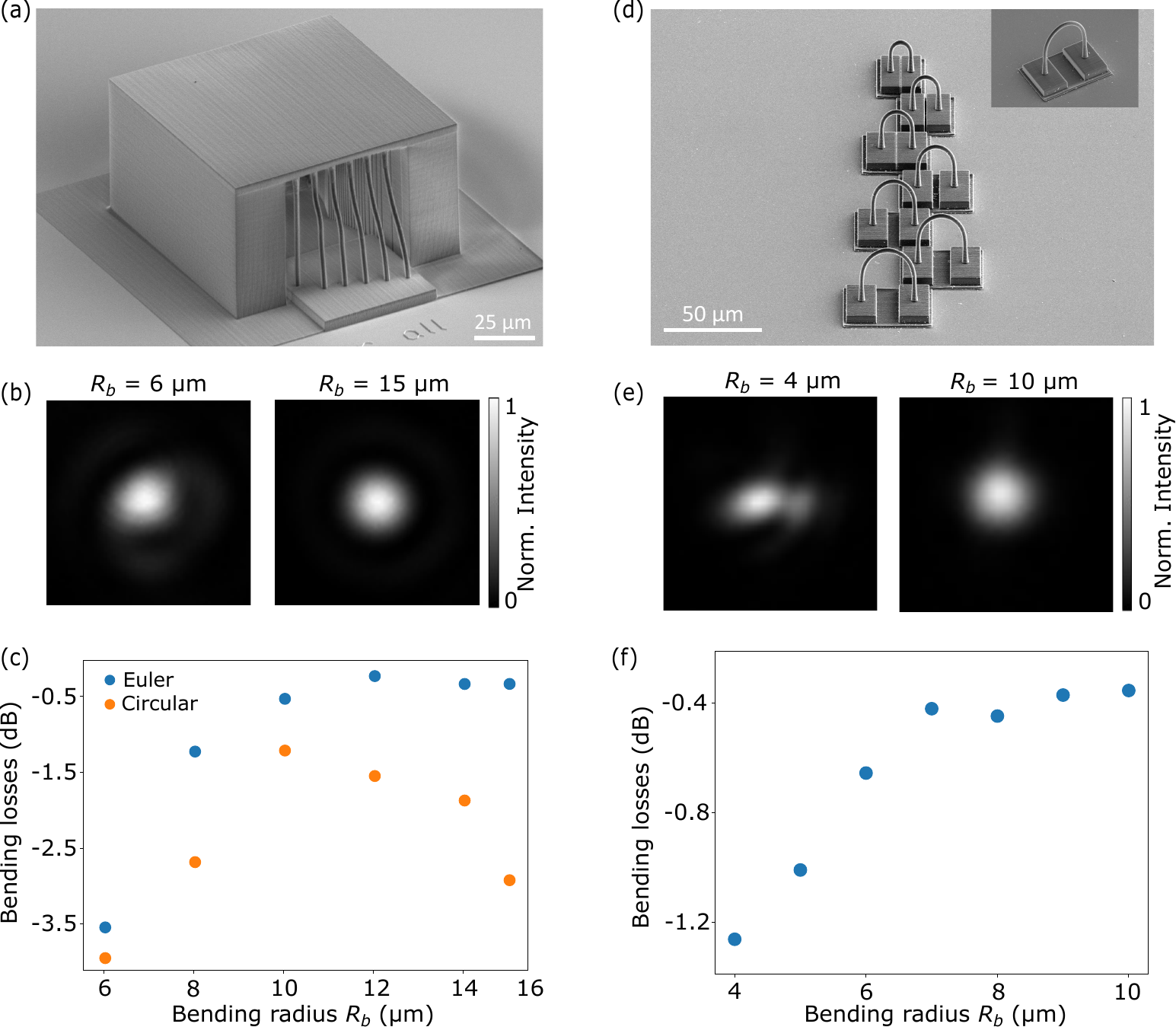}
\caption{(a) SEM micrograph of S-bends with varying bending radii. (b) Output intensity profile for $R_b = 6.0~\mu$m and $R_b = 15~\mu$m. (c) Bending losses for S-bends as a function of bending radii for Euler (blue) and circular geometry (orange). (d) SEM micrograph of U-bends with varying bending radii. (e) Output intensity profile for $R_b = 4.0~\mu$m and $R_b = 15~\mu$m. (f) Bending losses for U-bends as a function of bending radii.}
\label{fig:figure3}
\end{figure*}

\section{Air and polymer-clad waveguides}

\begin{figure*}[ht]
\centering\includegraphics[width=0.75\linewidth]{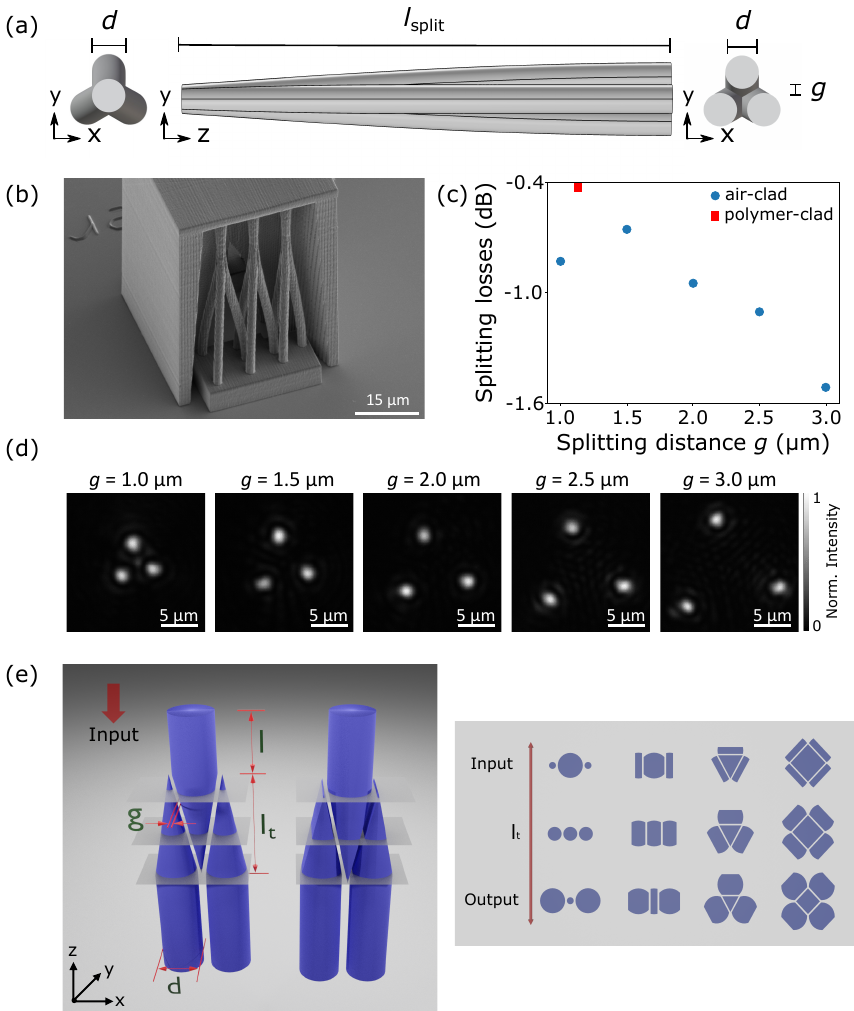}
\caption{(a) Schematic of air-clad splitters. (b) SEM micrograph of air-clad splitters. (c) Losses vs splitting distance for air-clad splitter (blue) and polymer-clad splitter (red, extracted from \cite{Grabulosa2023a}). (d) Output intensity profile for air-clad splitters for different gap $g$. (e) Schematic of polymer-clad splitter (left panel) and schematic of top view at different planes; input, transition, and output (right panel), reproduced with permission from \cite{Grabulosa2023a}.}
\label{fig:figure4}
\end{figure*}

Waveguides are fundamental building blocks of integrated optics, enabling a wide range of photonic circuits, including optical interconnects, modulators, lasers, and sensors. In this section, we will elaborate on the specifications and the need for merging both air-clad (high-confinement) and polymer-clad (low-confinement) waveguides into one integration platform, where the optical confinement is governed by the refractive index contrast between the core and cladding, defined as $\Delta{n}=n_{\textrm{co}} - n_{\textrm{cl}}$.
 Both waveguide types are fabricated in the same single step with our (3+1)D process where the degree of polymerization is controlled on a single voxel-level \cite{Grabulosa2022, Porte2021}. Furthermore, we leveraged flash-TPP, a 3D printing methodology combining TPP and one-photon polymerization (OPP). In an initial fabrication step, the waveguide cores of both, air-clad and polymer-clad sections, are fabricated via TPP as a 3D free-standing cylindrical pillar. For the air-clad waveguides the surrounding resin is removed during development, while for the polymer-clad waveguide the core remains immersed within unpolymerized resin encapsulated in a TPP-printed cage. For polymer-clad waveguide, the structure is then exposed to UV blanket irradiation ($3000~\mathrm{mJ/cm^2}$) to polymerize the cladding using OPP \cite{Grabulosa2022}. The schematic of air-clad and polymer-clad waveguide types is shown in Figure~\ref{fig:figure1}(a) and Figure~\ref{fig:figure1}(b), respectively.

\subsection{Optical confinement}

The 3D printed polymer-clad waveguides fabricated via flash-TPP have a refractive contrast $\Delta n\approx 6 \times 10^{-3}$ between core and cladding, and for $\lambda=650~$nm this results in a single-mode cut-off at a waveguide radius of $2.1~\mu$m. In contrast, the air-clad waveguide has $\Delta n\approx 0.5$, resulting in a single-mode cut-off at a waveguide radius of only $0.22~\mu$m. Figure 1(c) shows the effective refractive index ($n_{\mathrm{eff}}$) of the fundamental ($\mathrm{LP}_{01}$) and first higher-order ($\mathrm{LP}_{11}$) modes for air-clad and polymer-clad waveguides as a function of core radius, highlighting the modal evolution with increasing waveguide size, obtained using a finite-element eigenmode solver in COMSOL Multiphysics. Considering the $0.6~\mu$m voxel width for the used resin (IP-S) and objective ($25\times$) means that air-clad waveguides are multimodal at the minimum TPP feature size and have high surface roughness, while single-mode polymer-clad waveguides can readily be fabricated with high surface quality.

\subsection{Losses vs. bending radius}
\label{sec:Losses vs bending radius}

Compared to standard straight waveguides, the $n\textsubscript{eff}$ of the curved waveguides is spatially dependent on the bending radius $R_b$. When propagating in a curved waveguide, optical modes are displaced towards the cladding, where they attenuate exponentially. This leads to intermodal cross-talk from the fundamental mode into higher-order modes which consequently affects the propagation losses and the mode profile. Bending losses of circular bending waveguides are determined by the
following equation \cite{Lapointe2020,SnyderLove1983}:

\begin{align}
L_b = 1000 \times \left( 
\frac{2.171 \sqrt{\pi}}{\sqrt{RR_b}} \cdot 
\frac{v^4}{(v+1)^2 \sqrt{v-1}} \cdot \right. \nonumber \\
\left. \exp\left[ \frac{(v-1)^2}{v+1} - 
\frac{4 R_b (v-1)^3}{3Rv^2} \cdot 
\frac{n_{\text{co}}^2 - n_{\text{cl}}^2}{2 n_{\text{co}}^2} \right]
\right),
\label{eq:bend_loss}
\end{align}

\noindent where $v = \frac{2 \pi R}{\lambda} \sqrt{n_{\text{co}}^{2} - n_{\text{cl}}^{2}} $ is the V-value or the normalized frequency normalized frequency, $\lambda$ is the operating wavelength,  $R$ is the waveguide core radius and $R_b$ is the bend radius.

\begin{figure}[ht]
\centering\includegraphics[width=\linewidth]{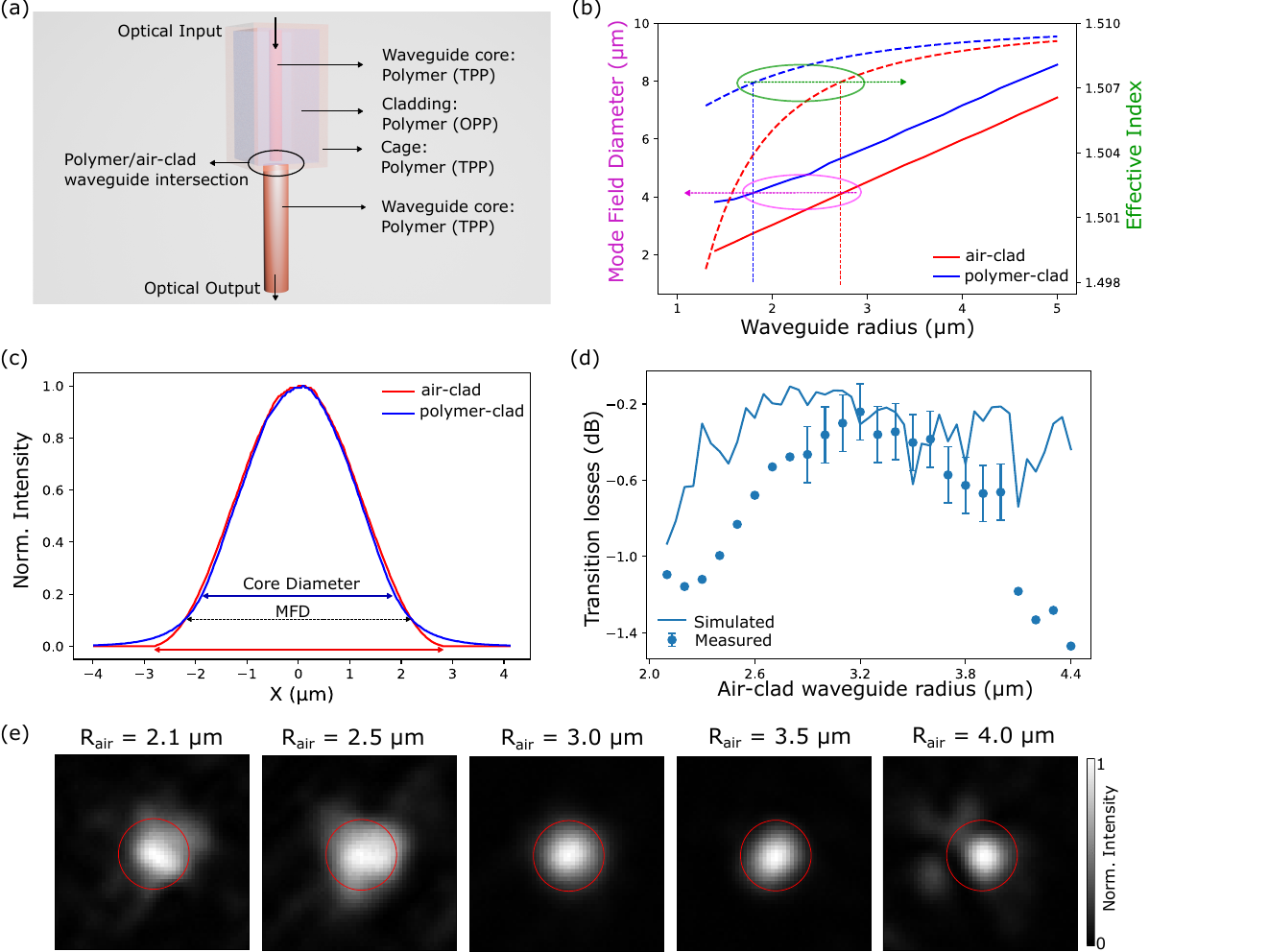}
\caption{(a) Schematic of a hybrid waveguide. (b) Mode field diameter (Effective index) for the {LP\textsubscript{01}} mode in air-clad (red) and polymer-clad (blue) waveguide on the left (right) y-axis as a function of the waveguide radius. (c) Gaussian profile for the air-clad (red) and polymer-clad (blue) waveguide; the solid line and dotted arrow line represent the waveguide core and MFD, respectively. (d) Transition losses with an error bar plot for the low losses range and as a function of radius of the air-clad waveguide along with the simulated losses. (e) Output intensity profile of hybrid waveguides at different radii with red circle indicating the waveguide core at the output.}
\label{fig:figure5}
\end{figure}

The dramatic difference in feasible bending radii for air and polymer-clad waveguides can be appreciated from the loss calculations shown in Figure~\ref{fig:figure1}(d). The higher $\Delta n$ of air-clad waveguides results in strong confinement of modes and supports tight bending for dense integration compared to polymer-clad waveguides. According to Eq.~\eqref{eq:bend_loss}, air-clad waveguides exhibits $0.01$ dB bending losses with $R_b = 10~\mu \textrm{m}$, while in contrast the polymer-clad waveguide requires $R_b = 1~\textrm{mm}$ to achieve the same losses. This difference in terms of minimal-bending radii for the two integration approaches is indeed significant for realizing high-density polymer-based photonic integrated circuits.

\subsubsection{Design and fabrication}
To design a low-loss curved waveguides, various bending geometries such as simple cosine or sine-like shapes, classical circular bends and Euler or Bezier curves have been studied \cite{Jiang2018, Farin1983}. For circular bends the curvature is constant over the whole bending process whereas for Euler bends the curvature is linearly decreasing. Figure~\ref{fig:figure2}(a) illustrates the geometrical shapes for both Euler (blue) and circular (red) geometries, illustrating how an Euler bend starts with a straight section with $R_b = \infty$ from where on the curvature radius reduces until it reaches $R_b = R\textsubscript{min}$. Figure~\ref{fig:figure2}(b) illustrates the $90^\circ$ Euler and a circular bend with bending radius $R_b = 1$ as the constant (circular) and minimum bending radius (Euler). For a symmetrical Euler-bend of angle $\alpha$ interfaced with straight in and output waveguides, two identical clothoid of angle $\alpha/2$ are combined. Under Euler bend conditions, the mode's location adiabatically approaches the core-cladding interface until reaching its maximally displaced location at $R_b=R_\text{min}$. This is contrary to other bending strategies, where such an adiabatic transition between straight and bend sections, and consequently adiabatic mode displacement from the waveguide center, is not guaranteed.

Here, the curved waveguides and splitters are fabricated via DLW-TPP with commercially available photoresist IP-Dip and a $63\times$ objective with an NA of 1.4, on a fused silica substrate using the commercial tool from Nanoscribe GmbH (Photonic GT+). Before the TPP fabrication, the fused silica is activated by oxygen plasma treatment for improved photoresist adhesion. Bends are than fabricated using TPP with optimal printing laser power ($\textrm{LP} = 9~\textrm{mW}$) and minimal hatching distance ($h = 0.1\mu \textrm{m}$), defined as the lateral spacing between adjacent scan lines, for smooth interfaces. A mechanical support in a form of semi-open cage structure is designed to support the air-clad waveguides to ensure their stability during and after the development procedure. Mechanical supports are printed with $\textrm{LP} = 15~\textrm{mW}$ and $h = 0.4 ~\mu\textrm{m} $ for reduced printing time\cite{Grabulosa2022}. A mechanical pedestal of height = $0.4~\mu\textrm{m}$ is designed for the increased adhesion of the curved waveguides to the glass substrate with an initial layer of larger TPP-dosage ($\textrm{LP} = 10~\textrm{mW}$) followed by the usual printing dosage used for the bends, see Figure~\ref{fig:figure2}(c). The slicing distance is kept fixed at $s = 0.1~\mu\mathrm{m}$ for the complete structure, where $s$ denotes the vertical spacing between consecutive layers during fabrication. This parameter significantly influences the surface roughness and structural stability of the curved waveguides. For the optical characterization, throughout section \ref{sec:Losses vs bending radius}, bending losses are measured by referencing to the global losses in a straight waveguide.

\subsubsection{S-shaped and U-shaped bends}
First, we investigate S-bends based on Euler and circular geometries, each consisting of two bending angles with $\alpha = 45^\circ$ and opposite curvature, resulting in a $180^\circ$ rotation of bending orientation between successive sections. In the Euler configuration, each $45^\circ$ bend is composed of two consecutive $22.5^\circ$ Euler segments to ensure continuous curvature transition. Such S-bends are commonly used to connect two points offset by a distance, for example, coupling adjacent interferometers in 2D meshes of Mach-Zehnder \cite{Baghdasaryan2021} and fan-out sections in splitters. Bending losses are measured over the range of bending radii spanning \( R_b \in \{6, \dots, 15\}~\mu\textrm{m} \) and by removing injection and propagation losses. Figure~\ref{fig:figure3}(a) shows an SEM micrograph of such S-bends with varying radii and a straight waveguide serving as reference for injection and pure propagation losses. Figure~\ref{fig:figure3}(b) shows the output intensity profiles at $R_b = 6~\mu\textrm{m}$ and $R_b = 15~\mu\textrm{m}$, clearly showing that while for $R_b = 6 ~\mu\textrm{m}$ the waveguide's output is multi-mode, for $R_b \geq 10 ~\mu\textrm{m}$ the waveguide's output is single-mode and well centered. In Figure~\ref{fig:figure3}(c), we compare the experimentally determined bending losses for circular (orange) with the ones for Euler (blue) S-bends. The superior performance of Euler over circular S-bends is clearly apparent, with Euler bends achieving below 0.5 dB bending losses for bending radii of 10$~\mu\textrm{m}$ and larger. Notably, for S-bends, the losses increase for bend radii $R_b \geq 10 ~\mu\textrm{m}$ when using circular bends, which is potentially a consequence of fabrication non-idealities, including mechanical instability and variations in the waveguide cross-section along the length.

The next study of curved waveguides is done for U-bends with bending angle $\alpha = 180^\circ$, implemented using two consecutive $90^\circ$ Euler-bends. For this reason, and due to the clear advantage of Euler over circular bends demonstrated before, we here do not show results for the circular U-bends. Figure~\ref{fig:figure3}(d) shows the SEM of U-bends with bending radii of \( R_b \in \{4, \dots, 10\}~\mu\textrm{m} \). From the output intensity profile shown in Figure~\ref{fig:figure3}(e), it is clear that at a bending radius of $R_b = 4~\mu\textrm{m}$ higher modes are observed, while for $R_b = 10~\mu\textrm{m}$ the output is a single fundamental mode and centered at the core. Figure~\ref{fig:figure3}(f) shows the corresponding bending losses, which decreases until it reaches a plateau at $0.4~\textrm{dB}$ for $R_b > 7~\mu\textrm{m}$.

\section{Air-clad 3D splitters}
\label{sec:splitters}

Optical routing between multiple ports is an essential functionality for integrating 3D photonic circuits. In \cite{Grabulosa2023a}, high-quality 1-to-M broadband adiabatic couplers based on polymer-clad waveguides fabricated via flash-TPP were demonstrated. In this section, we report adiabatic air-clad splitters that due to the increased confinement features a significantly smaller footprint and low splitting losses compared to previously reported values in \cite{Baghdasaryan2024}. The layout of the air-clad splitters is shown in Figure~\ref{fig:figure4}(a) for a 1 to 3 air-clad splitters and Figure~\ref{fig:figure4}(b) shows the SEM micrograph of the same. The cross-section of each output waveguide is extruded from the input waveguide along the parametric path. For a splitter with $K$ output ports, where \(k \in \{0,1,...,K-1\}\) denotes an output branch, the coordinates of each branch are obtained through rotation about the z-axis:

\begin{align}
  \begin{pmatrix}
  x \\[4pt]
  y \\[4pt]
  z
  \end{pmatrix}
  =
  R_z\!\left(\tfrac{2\pi k}{K}\right)
  \begin{pmatrix}
  x_0(j) \\[4pt]
  y_0(j) \\[4pt]
  z_0(j)
  \end{pmatrix},
\end{align}
where the roation matrix is defined as
\begin{align}
  R_z\!\left(\tfrac{2\pi k}{K}\right)=
  \begin{pmatrix}
  \cos\!\left(\tfrac{2\pi k}{K}\right) & -\sin\!\left(\tfrac{2\pi k}{K}\right) & 0\\[6pt]
  \sin\!\left(\tfrac{2\pi k}{K}\right) &  \cos\!\left(\tfrac{2\pi k}{K}\right) & 0\\[6pt]
  0 & 0 & 1
  \end{pmatrix},
\end{align}
and

\begin{align}
\begin{pmatrix}
x_0(j)\\
y_0(j)\\
z_0(j)
\end{pmatrix}
=
\begin{pmatrix}
\frac{g}{2}\bigl(\cos(\pi j)-1\bigr)\\
0 \\
l_{\mathrm{split}}\,.j
\end{pmatrix}
\end{align}
describes the parametric curve of a single branch before rotation, where \(g\) is the splitting distance between the individual output ports, while \( j=z / l\textsubscript{split} \in \{0, \dots, 1\} \) is the index for slicing the parametric path in z-direction normalized by the splitting length (\(l\textsubscript{split}\)) during which the splitter's individual output ports gradually separate. 

For the optical characterization, splitting losses are measured by referencing to the global losses in a straight waveguide of the same height. Figure~\ref{fig:figure4}(c) shows the splitting losses over a range of separation gaps \(g \in \{1.0:0.5:3.0\}\)$~\mu$m, with global losses increasing for larger $g$. This is expected, as we keep the height of the overall structure and  $l\textsubscript{split} = 52~\mu$m constant for all $g$, and hence a larger $g$ leads to tighter bending radii. In addition to optical losses, this separation also significantly affects the system's mode profile at the 3-output ports. As shown in Figure~\ref{fig:figure4}(d), for $g = 1.0 ~\mu$m the output intensity profiles are not yet fully decoupled, with some residual optical intensity visible in the center between the three output waveguides. A complete splitting of the injected single-mode input into three fundamental output modes is achieved starting at separation gaps $g = 1.5~\mu$m, with negligible cross-talk and splitting losses of $0.6~\textrm{dB}$. 

Figure~\ref{fig:figure4}(e) illustrates the schematic of 1 to 2 polymer-clad splitters with the left panel showing a truncated taper cross-section geometry and its right panel a cross-section view at 3 different stages along the z-axis. This truncated geometry showed lower coupling losses compared to the conical geometry due to increased effective area coupling between the waveguides with coupling losses as low as $0.4$ dB for coupling taper length $l\textsubscript{t} = 500~\mu$m and gap $g = 0.4~\mu$m \cite{Grabulosa2023a}.

\begin{figure}[ht]
\centering\includegraphics[width=\linewidth]{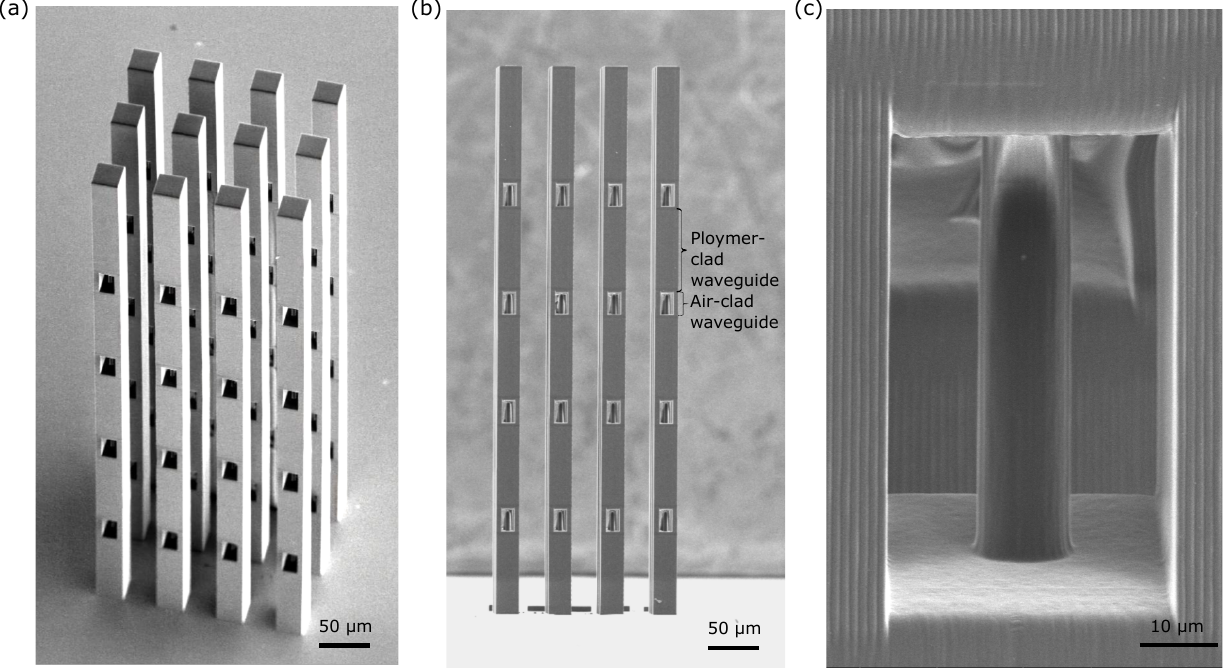}
\caption{(a) Side and (b) top view of SEM micrograph of the array of 3D printed hybrid waveguides. (c) SEM micrograph of the air-clad section in the hybrid waveguide.}
\label{fig:figure6}
\end{figure}

It is worth highlighting the individual aspects of the two approaches that are taken in this section in order to obtain splitting of a single-mode into various single-mode outputs. Our novel air-clad concept leverages the geometric modification of the guiding core which merges from the one input into the 3 outputs without interruption. The reason for this approach lies in the strong confinement and the relatively low printing resolution, prohibiting the usage of an evanescent leakage  field for distributing an input across various outputs. It therefore uses the confined instead of the evanescent field, and hence we obtain significantly larger coupling rates between the splitter's output ports, while maintaining their single-mode nature by carefully engineering the overall design. As a consequence, the dimensions of air-clad splitters have been reduced to a length of tens of micrometers ($l\textsubscript{split} = 52~\mu$m) compared to hundreds of micrometers ($l\textsubscript{t} = 500~\mu$m) for polymer-clad while obtaining comparable coupling losses $0.6$ dB and $0.4$ dB, respectively. Thus, air-clad splitters and their non-evanescant coupling concepts provide a clear advantage over polymer-clad splitters with regards to increased component density for future 2D as well as 3D integration concepts. Finally, it is important to note that the 3D fabrication approach employed here is critical to the achieved performance compared to conventional 2D lithographic techniques, which typically offer only binary control over the confinement geometry, while 3D enables adiabatic sculpting of the mode in all three dimensions \cite{Nikkhah_2024, Brunner2024}.

\section{Hybrid Waveguide Circuits}

\subsection{Mode control and high-density integration using hybrid-confinement circuits}

As previously discussed, polymer-clad waveguides are single-mode in the resolution limit of 3D TPP printing. Their low and controllable confinement when leveraging (3+1)D \cite{Porte2021} and flash-TPP \cite{Grabulosa2022} makes them highly suitable for evanescent coupling and mode control in hybrid waveguide's core. Contrary, air-clad waveguides support tight bends as well as rapid single-mode fan-in and fan-out solutions due to strong confinement. Although their confinement is multi-mode within the resolution limit of the DLW-TPP fabrication process, single-mode operation of the propagating mode can be maintained if the injected mode overlaps with the fundamental mode of the multi-mode confinement section with high fidelity \cite{SnyderLove1983,Yariv1973}. This single-mode operation of a multi-mode circuit is, for example, a strategy for boosting the quality factor in microrings, where the reduced evanescent field of the fundamental mode in a multi-mode waveguide strongly reduces scattering losses at the interface of the guiding and the cladding element \cite{OptCom2015LowLoss}.

This suggests a novel integration method with DLW-TPP, which allows highly matched fundamental modes for polymer-clad and air-clad waveguides with a smooth fundamental mode propagation throughout the circuit. To leverage the best of the air-clad and polymer-clad 3D waveguide worlds, we have designed hybrid waveguide circuits that combine polymer-clad and air-clad waveguides with various degrees of polymer/air-clad intersections, schematically illustrated in Figure~\ref{fig:figure5}(a). The abrupt transmission of a mode from the weakly confined polymer-clad waveguide to the strongly confined air-clad waveguide requires accurate mode overlap and matching of the $n\textsubscript{eff}$ to avoid significant back reflections at such confinement transitions.

Transition losses at the interface between the two confinement sections can be calculated via the overlap coupling efficiency of the individual electric field profiles of the modes before ($E_1$) and after ($E_2$) the confinement transition according to \cite{SalehTeich2007, SnyderLove1983}:
\begin{align} 
\eta = \frac{\left| \iint E_1(x,y) \, E_2^*(x,y) \, dx\,dy \right|^2}
{\iint |E_1(x,y)|^2 \, dx\,dy \; \iint |E_2(x,y)|^2 \, dx\,dy}.
\end{align}
The analytical solution for the above coupling efficiency can be simplified when assuming Gaussian profiles in the polymer-clad as well as air-clad waveguides based on mode field diameters (MFDs) \cite{SalehTeich2007, SnyderLove1983}:
\begin{align}  
\eta = \left(\frac{2\,w\textsubscript{air} w\textsubscript{pol}}{w\textsubscript{air}^2 + w\textsubscript{pol}^2}\right)^{2},
\end{align}
where $w$ denotes the mode field radius, defined as the radial distance at which the field amplitude decreases to $1/e^{2}$ of its peak value, such that $\mathrm{MFD} = 2w$. To achieve unity coupling efficiency, the MFD of both air-clad and polymer-clad waveguides' fundamental modes need to be matched ($w\textsubscript{air} = w\textsubscript{pol}$). The fundamental modes of both air-clad and polymer-clad waveguides were numerically simulated using the finite-element eigenmode solver in COMSOL Multiphysics with core refractive index (TPP) $n\textsubscript{co}=1.51$ (obtained from the manufacturer’s datasheet \cite{Nanoscribe}) and cladding refractive index (OPP) $n\textsubscript{cl} = 1.5039$ (extracted by fitting the experimentally obtained fundamental mode profile \cite{Grabulosa2022}), and the resulting MFDs and corresponding effective indices for the respective modes are plotted over a range of varying waveguide radii. For the polymer-clad waveguide, the waist of the fundamental mode at the core radius below the single-mode cut off $R\textsubscript{pol} = 1.8~\mu$m sets the reference MFD and $n\textsubscript{eff} $ for our hybrid circuit. In our numerical simulations, we find MFD\textsubscript{pol} = 4.2$~\mu$m ($n\textsubscript{eff} = 1.5074 $) for these conditions, which is matched to the waist and $n\textsubscript{eff}$ of air-clad waveguide's fundamental mode at $R\textsubscript{air} = 2.8~\mu$m, see Figure~\ref{fig:figure5}(b). The corresponding 1D Gaussian intensity profiles of both the waveguides' fundamental modes indicate almost perfectly matched modes, see Figure~\ref{fig:figure5}(c), and accordingly the numerically obtained coupling efficiency $\eta$ for these parameters is unity.

\subsection{Design, fabrication and transition losses in hybrid waveguide circuits}
\label{subsec:fabrication}

Flash TPP and (3+1)D fabrication were used to fabricate the integrated hybrid waveguide circuits. For the fabrication of hybrid waveguides, IP-S with a $25\times$ microscope objective (NA = 0.8) was used instead of IP-Dip as mentioned in the earlier section. Compared to IP-Dip, IP-S has a faster printing speed, smooth surfaces, high mechanical endurance, and does not require pre-oxygen treatment of the glass substrate to enhance substrate adhesion, which makes it well-suited to print large and scalable structures. Furthermore, our studies found that IP-Dip polymerizes too rapidly for the optical fluences available in our UV-chamber to allow efficient and controlled flash-TPP fabrication, though this is not a fundamental challenge and IP-Dip based hybrid 3D integration should be possible in principle.

The strongly confined air-clad waveguide core is printed with TPP using a hatching distance of $h=0.1~\mu\textrm{m}$, slicing distance of $s=1~\mu\textrm{m}$, and laser power of LP = 20 mW. While the weakly confined polymer-clad waveguide is printed by flash-TPP \cite{Grabulosa2022} with it's core printed with $h=0.4 ~\mu \textrm{m}, ~s=1~\mu \textrm{m}$ and LP = 37 mW and for the cage higher hatching distance of $0.8 ~\mu \textrm{m}$ with $~s=1~\mu \textrm{m}$ and LP = 45 mW was used to have reduced printing time. The lower hatching distance was chosen for the air-clad core in order to fine tune the radii, and subsequently lower LP was used to avoid the burning of the resin caused by over-polymerization. To accurately determine the excess losses at these confinement transitions, we 3D print structures in which we alternate between strongly and weakly confined waveguide sections.

\begin{figure}[ht]
\centering\includegraphics[width=\linewidth]{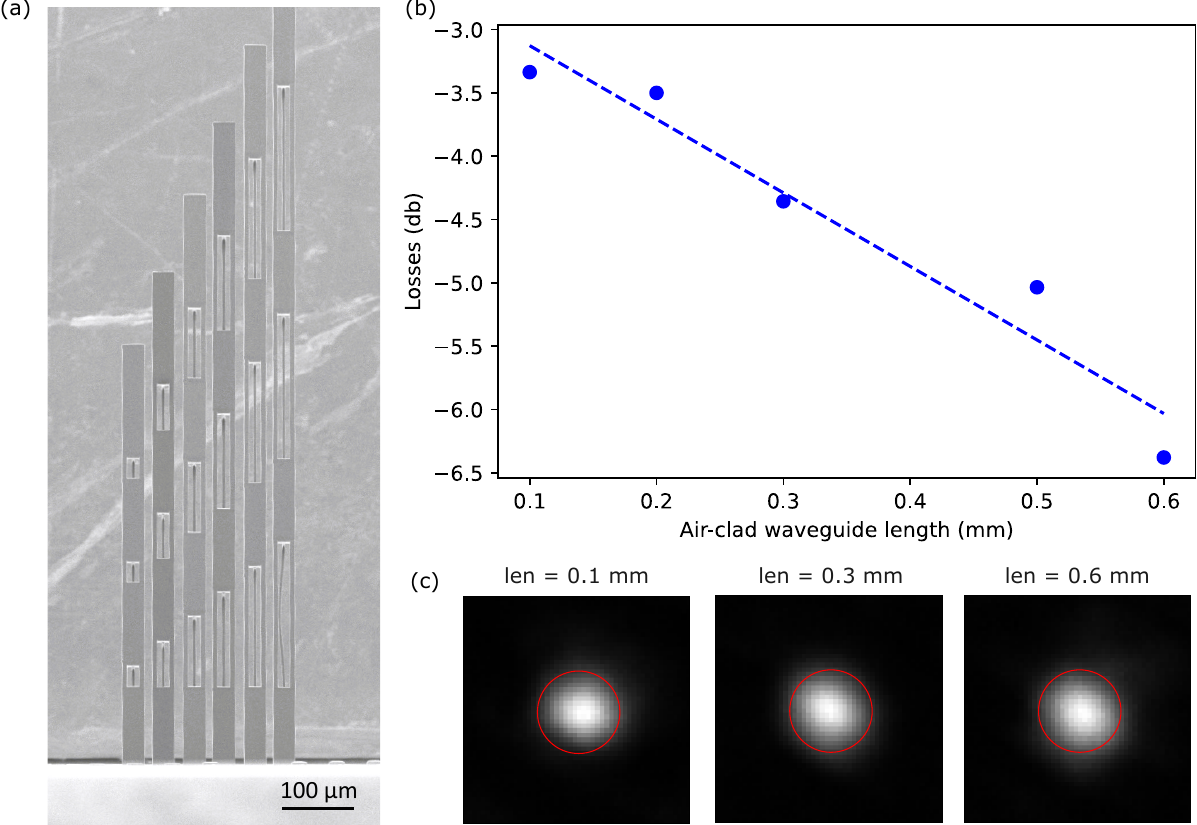}
\caption{(a) SEM micrograph of 3D printed air-clad waveguide sandwiched between the polymer-clad waveguide. (b) Plot of losses in an air-clad waveguide as a function of length with a linear fit. (c) Output intensity profile of hybrid waveguides for different lengths of air-clad waveguide with red circle indicating the waveguide core at the output.}
\label{fig:figure7}
\end{figure}

Figure~\ref{fig:figure6} shows the SEM micrograph for such hybrid waveguides with constant $R\textsubscript{pol} = 1.8~\mu$m, while the radius of the air-clad waveguide was swept \( R\textsubscript{air} \in \{2.1, \dots, 4.4\}~\mu \textrm{m} \) to characterize the fundamental mode's transmission efficiency in terms of losses as well as maintaining the fundamental mode character through the circuit. The structure was cascaded with four poly-air-poly transition cells, yielding a total of 8 confinement transitions. Figure~\ref{fig:figure5}(d) shows the measured optical transition losses of the hybrid waveguide together with numerical simulations performed using the 2D Electromagnetic Wave Equation, Frequency Domain study in COMSOL Multiphysics. The lowest confinement transition loss obtained is 0.25 dB for $R_{\mathrm{air}} = 3.2~\mu\mathrm{m}$, in good agreement with the numerical predictions. The discrepancy between the experimental and simulated results beyond $R_{\mathrm{air}} = 3.8~\mu\mathrm{m}$ can be attributed to deviations from an ideal circular waveguide geometry arising from the finite hatching distance during fabrication, which affects the optical mode profile differently depending on the degree of mode confinement. To calculate the confinement transition losses, we subtracted the measured injection losses ($0.06$ dB), propagation losses in polymer-clad waveguide (1.39 dB/mm), which we obtained from individual reference measurements that are also in accordance with previously published results \cite{Grabulosa2022}, and propagation losses in air-clad waveguide (measured in Section~\ref{sec:air-clad loss}) from the global losses. The resulting confinement transition loss is as low as 0.25 dB at $R\textsubscript{air} = 3.2~\mu$m and $R\textsubscript{pol} = 1.8~\mu$m and showing an excellent agreement with the numerical simulations in terms of losses as well as optimal core-radii. 

The optical output mode after traversing 4 consecutive transitions between air/polymer-clad cells, i.e., after traversing 8 confinement transitions, is shown as the output-waveguide's optical near field in Figure~\ref{fig:figure5}(e). We show the near fields for air-clad waveguide radii $R\textsubscript{air}$ = 2.1, 2.5, 3.0, 3.5, and 4.0 $~\mu$m, with the red circle indicating the output polymer-clad waveguide circumference. Deviating away significantly shows the increased optical scattering as well as in a speckled multimodal output field. The intensity profile clearly shows that for $R = 3.0~\mu$m the output intensity is centered within the waveguide and excellently follows the Gaussian profile of a fundamental mode. The detailed mapping of transition efficiency as well as its agreement of fundamental output mode at the points of maximal coupling efficiency is a robust confirmation of our approach.

\subsection{Propagation losses in air-clad waveguide}
\label{sec:air-clad loss}

\begin{figure*}[ht]
\centering\includegraphics[width=0.75\textwidth]{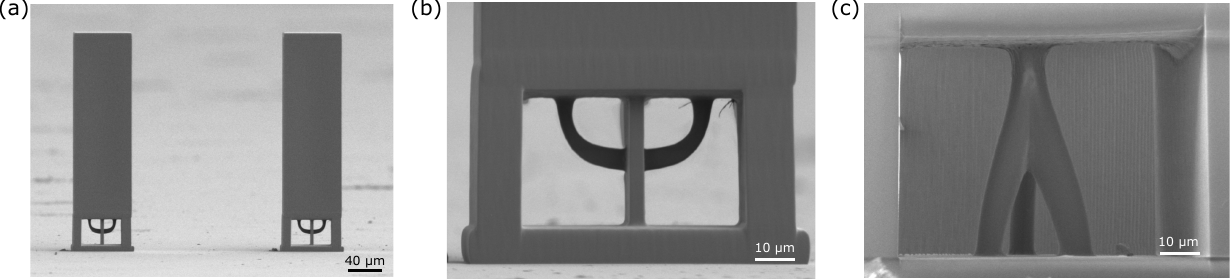}
\caption{SEM micrograph of fabricated (a) 3D hybrid waveguide circuit, (b) an air-clad U-bend connected to polymer-clad waveguides at both input and output, and (c) air-clad splitter preceded by the polymer-clad input waveguide.}
\label{fig:figure8}
\end{figure*}

Beyond the previous fundamental characterization, the hybrid waveguide circuit paved the way for measuring the propagation losses of the fundamental mode $\mathrm{LP}_{01} $ in the air-clad waveguide. The incorporation of a well-designed polymer-clad section enables efficient and robust coupling of the fundamental mode into the multi-mode air-clad waveguide, which would otherwise be challenging due to mode mismatch at the interface. In addition, the polymer-clad sections enhance structural robustness and provide improved control over modal evolution. This hybrid design allows the realization of significantly longer air-clad propagation sections, sandwiched between polymer-clad regions. To experimentally quantify the propagation loss of the fundamental mode in the air-clad waveguides, we fabricated hybrid circuits with varying air-clad lengths \( l\textsubscript{air} \in \{100, 200, \dots, 600\}~\mu \textrm{m} \), as shown in the SEM micrographs in Figure~\ref{fig:figure7}(a). The mode profile and global losses of hybrid waveguides were measured as a function of the air-clad waveguide length. Consistent with our previous results, we observe single-mode operation at $R\textsubscript{air} = 3.0~\mu$m for different lengths of air-clad waveguide, see Figure~\ref{fig:figure7}(c). The air-clad propagation loss was determined from the slope of a linear fit to the measured global loss of the hybrid waveguide circuit as a function of air-clad waveguide length. From the fit to our experimental data, see Figure~\ref{fig:figure7}(b), we obtain propagation losses of the mode $\mathrm{LP}_{01} $ in air-clad waveguides as $5.8$ dB/mm at $R\textsubscript{air} = 3.0~\mu$m.

\subsection{3D Integrated Optical Devices}

In this section, we present a more advanced functionality consisting of a polymer-clad straight waveguide followed by compact air-clad splitters and U-bends, all fabricated using IP-S, see Figure~\ref{fig:figure8}. To ensure smooth splitter and bend profiles, hatching distance of $h=0.1~\mu\textrm{m}$, slicing distance of $s=0.1~\mu\textrm{m}$, and laser power of LP = 10 mW were used, while for polymer-clad section the same printing parameters were used as mentioned in Section~\ref{subsec:fabrication}. The measured losses for the air-clad splitter and U-bend are 1.5 dB and 2 dB, respectively, with the single mode propagation maintained throughout the circuit. This degradation in losses is primarily attributed to voxel asymmetry during fabrication. In particular, the theoretical axial resolution is approximately $0.7~\mu\mathrm{m}$ for the 63× objective and $3.6~\mu\mathrm{m}$ for the 25× objective \cite{Nanoscribe}, limiting the accurate fabrication of bends and splitters with waveguide radius of $1~\mu\mathrm{m}$ using IP-S. These limitations can be mitigated by operating at longer wavelengths, such as the telecommunications band (1550 nm), rather than the current 650 nm, or by using (3+1)D and flash-TPP with the IP-Dip resin.

\bigskip

\section{Discussion and conclusion}
In summary, we successfully demonstrated 3D hybrid-confinement photonic circuits using a commercially available DLW-TPP system from Nanoscribe GmbH. High-confinement air-clad waveguides are employed to realize tight bends and compact splitters. S-shaped and U-shaped bends based on Euler geometry exhibit bending losses of 0.5 dB and 0.4 dB, respectively, for a bending radius of $10~\mu\mathrm{m}$. Air-clad splitters achieve a splitting loss of $0.6~\mathrm{dB}$ over a compact length of $52~\mu\mathrm{m}$, corresponding to a tenfold reduction in printing volume compared to previously reported evanescent-field-based splitters \cite{Grabulosa2023a} and up to fivefold reduction in splitting losses reported in air-clad splitters from \cite{Baghdasaryan2024}. The bends and splitters are fabricated using the IP-Dip photoresist and $63\times$ objective.

We then introduced the hybrid waveguide circuit for smooth single-mode manipulation by transitioning from a single-mode polymer-clad waveguide to a multi-mode air-clad waveguide, achieving transition losses of 0.25 dB. Using this hybrid technology, the propagation loss of $5.8~\textrm{dB/mm}$ for mode $\mathrm{LP}_{01} $ is measured in air-clad waveguides. The hybrid circuits are fabricated using the IP-S photoresist with a $25\times$ objective, owing to its higher printing speed and the absence of pre-oxygen treatment requirements compared to IP-Dip. In addition, the minimum controllable UV exposure dose of our system ($750~\mathrm{mJ/cm^2}$) is sufficient to fully polymerize IP-Dip, thereby precluding its use in the high-resolution flash-TPP configuration at this stage. Finally, full circuit integration is demonstrated on a single chip with a single resin IP-S, with slightly increased losses of bends (2 dB) and splitters (1.5 dB) attributed to voxel asymmetry and axial resolution limitations ($3.6~\mu\textrm{m}$), which constrain the fabrication of $2~\mu\textrm{m}$ features.

The proposed hybrid waveguide circuit demonstrate a significant step towards large-scale photonic 3D integration leveraging the best of both regimes; air-clad waveguides can be used for tight bends and dense integration, while polymer-clad waveguide can be used for the mode manipulation and circuit sections requiring leveraging evanescent fields. The approach provides a scalable pathway toward applications such as photonic wire bonding, 3D optical neural networks, and photonic lanterns. Further reduction in bending and splitting losses by more than a factor of four is anticipated through using flash-TPP technology with IP-Dip and a $63\times$ objective, enabling higher-resolution fabrication. In addition, operation at longer wavelengths, such as in the telecommunication band, would permit larger waveguide diameters in the air-clad sections within the current architecture, thereby reducing bending and splitting losses. Furthermore, longer-wavelength operation is expected to reduce both propagation and transition losses by mitigating surface and bulk Rayleigh scattering originating from the submicron voxelized structure.  
%%%%%%%%%%%%%%%%%%%%%%% References %%%%%%%%%%%%%%%%%%%%%%%%%

\bigskip

\noindent\textbf{Funding.} Project INSPIRE (HORIZON.1.1 - European Research Council (ERC)); French National Research Agency (ANR-22-PEEL-0010); Project Photonic QRC (No. Gi1121/6-1); Swiss National Science Foundation project LION.

\bigskip

\noindent\textbf{Acknowledgement.} The authors acknowledge the support of the
French RENATECH network and its FEMTO-ST technological facility MIMENTO. The authors also thank Gwenn Ulliac and Marina Raschetti for techincal support.

\bigskip

\noindent\textbf{Disclosures.} The authors declare no conflicts of interest.

%%%%%%%%%% If using BibTeX:
\bibliography{sample}
	
\end{document}